\documentclass[aps,prd,twocolumn,preprintnumbers,superscriptaddress,floatfix]{revtex4}

\usepackage{multirow, graphicx,amssymb,url,mathrsfs,amsmath}
\usepackage{eucal,wrapfig,boxedminipage,setspace,subfigure}
\usepackage{amsxtra,amstext,latexsym,dsfont}
\usepackage{xcolor}

\def\IR{{\hbox{{\rm I}\kern-.2em\hbox{\rm R}}}}
\def\IB{{\hbox{{\rm I}\kern-.2em\hbox{\rm B}}}}
\def\IN{{\hbox{{\rm I}\kern-.2em\hbox{\rm N}}}}
\def\IC{\,\,{\hbox{{\rm I}\kern-.59em\hbox{\bf C}}}}
\def\IZ{{\hbox{{\rm Z}\kern-.4em\hbox{\rm Z}}}}
\def\IP{{\hbox{{\rm I}\kern-.2em\hbox{\rm P}}}}
\def\IH{{\hbox{{\rm I}\kern-.4em\hbox{\rm H}}}}
\def\ID{{\hbox{{\rm I}\kern-.2em\hbox{\rm D}}}}

\newcommand{\beq}{\begin{equation}}
\newcommand{\eeq}{\end{equation}}
\newcommand{\bea}{\begin{eqnarray}}
\newcommand{\eea}{\end{eqnarray}}

\begin{document}

\voffset 1cm

\newcommand\sect[1]{\emph{#1}---}

\title{Holographic Colour Superconductors at Finite Coupling with NJL Interactions}

\author{Kazem Bitaghsir Fadafan}
\affiliation{ Faculty of Physics, Shahrood University of Technology,
P.O.Box 3619995161 Shahrood, Iran}

\author{Jes\'us Cruz Rojas}
\affiliation{ STAG Research Centre \&  Physics and Astronomy, University of
Southampton, Southampton, SO17 1BJ, UK}

\begin{abstract}

We study a bottom-up holographic description of the QCD colour superconducting phase in the presence of higher derivative corrections.  We expand this holographic model in the context of Gauss-Bonnet (GB) gravity. The Cooper pair condensate has been investigated in the deconfinement phase for different values of the GB coupling parameter $\lambda_{G B}$, we observe a change in the value of the critical chemical potential $\mu_c$ in comparison to Einstein gravity. We find that $\mu_c$ grows as $\lambda_{G B}$ increases. We add four fermion interactions and show that in
the presence of these corrections the main
interesting features of the model are still present and that the intrinsic 
attractive interaction can not be switched off. This study suggests to find GB corrections to equation of state of holographic QCD matter.
\end{abstract}%

\maketitle

\newpage
\section{Introduction}

It is widely known that in a fermionic system at finite chemical potential, where a Fermi surface is expected to develop, if there is any attractive interaction between the fermions, Cooper pair condensation will occur causing superconductivity or superfluidity. This fact leads to the natural expectation that quarks will condense at high density in quantum chromodynamics (QCD), and there has been considerable work on understanding the phase structure over the years (see the review \cite{Alford:2007xm}). 

Typically the preferred condensation pattern is expected to break the colour gauge group so the phenomena has been named colour superconductivity (CSC). An exact computation of the condensation can be done at very large chemical potential where QCD, because of asymptotic freedom, becomes weakly coupled \cite{Son:1998uk}. However, the more experimentally interesting case is when the density and temperature of the quark-gluon plasma (QGP) are of order the strong coupling scale $\Lambda_{c}$ where the strongly coupled nature of the problem prevents a precise computation. Here, gap equation and renormalization group study have been done and the possible phase structure of the system as a function of number of flavours, $N_f$ and the quark mass has been found \cite{Alford:2007xm}. First principle computations in QCD are currently inaccessible at this regime where temperature is low and chemical potential is high. The reason comes from strongly coupled nature of the system and the sign problem of lattice gauge theory \cite{Aarts:2015tyj}. 

Holography is a new tool to study strongly coupled gauge theories \cite{Maldacena:1997re}. It is using in ${\cal N}=4$ super Yang-Mills theories at large number of colours, $N_c$ and maps the computations to the weakly coupled gravitational theories. The quark degrees of freedom has been added in the gauge theories in \cite{Karch:2002sh} so that one can study wider space of such theories using $AdS/QCD$ \cite{Erlich:2005qh}. Holographic study of cool baryon density and quark matter \cite{Ishii:2019gta}, possible hadron-quark continuity \cite{BitaghsirFadafan:2018uzs} and quarkyonic matter \cite{Kovensky:2020xif} are such examples of applied holography in dense QCD. Although there is no large $N_c$ limit and CSC is sub-leading in this limit  \cite{Shuster:1999tn}, one expects an instability to pair condensation in the presence of chemical potential. Then a colour singlet pair can be formed \cite{Evans:2001ab} and used to the study of AdS superconductor phenomena in condensed matter, leading to the so-called AdS/CMT \cite{Hartnoll:2008vx}.

Study of the CSC phase of QCD using this new tool has been done in \cite{BitaghsirFadafan:2018iqr}, recently. Also, first principle holographic colour superconductivity with its spectrum have been studied in \cite{Faedo:2018fjw} and \cite{Faedo:2019jlp}. Furthermore, CSC considering backreaction has been done in \cite{Ghoroku:2019trx}. Another study has found a novel CSC phase in super Yang-Mills theory \cite{Henriksson:2019zph}. Recently CSC in Einstein-Gauss-Bonnet gravity has been investigated in the confined and  the deconfined phases \cite{Nam:2021qwv}. 

In this paper, we study the deconfined phase where the QGP, at intermediate density, is strongly coupled and below of the chiral phase transition. Here one expects that at the confiment-deconfiment transition magnetically charged scalars condense to cause confinement \cite{Ramamurti:2018evz}. However above of the transition such states generate a Debye mass for electric and magnetic gluons. Thus there would be a gap between the gluon mass and the chemical potential. Therefore this would allow us to apply holography in such system by a bottom up set up \cite{BitaghsirFadafan:2018iqr}. As an important application of this study, see \cite{BitaghsirFadafan:2020otb} where equation of state of colour superconductivity has been found and as a result it was deduced that quark matter could be present at the core of compact stars.

Here we want to expand the model proposed in \cite{BitaghsirFadafan:2018iqr} by considering an Einstein-Gauss-Bonnet space-time. The goal is to study the role of higher curvature corrections written as the Gauss-Bonnet (GB) term and study how the critical chemical potential at which the Cooper condensate is formed could be depended on this term. Holographic condensed matter superconductors in the presence of higher derivative corrections have been studied in \cite{Gregory:2009fj}. Following \cite{Ghoroku:2019trx}, the upper bound on $N_c$ has been recently computed in the presence of GB corrections in \cite{Nam:2021qwv} by considering backreaction on the background. However, as in \cite{BitaghsirFadafan:2018iqr}, we work in the probe limit and neglect the back reaction of the matter part to the geometry.

 We will also add four fermion interaction to study condensation in the presence of this new parameter. We know that the precise boundary dual theory to GB background is unknown but using the duality one can study the expectation value of operators in the dual theory. Then from the boundary field theory point of view, we are going to study the effect of finite coupling corrections on the holographic CSC phase of QCD which helps to construct the holographic model more accurately.

This paper is organized in the following way: in Section II we review the gravitational model dual to the CSC phase using the action of Einstein-Gauss-Bonnet gravity in five space-time. In Section III we  introduce NJL operators in the model. In section IV we study how the phase diagram changes with  Gauss-Bonnet corrections. We discuss the effect of higher derivative corrections in the last section.
\section{ CSC in Gauss Bonnet background}
We consider the action of Einstein-Gauss-Bonnet gravity in five space-time dimensions given by
\begin{equation} \label{GB_action}  \begin{array}{l}
S_{G B}=\frac{1}{2 \kappa_{5}^{2}} \int d^{5} x \sqrt{-g}[R+\frac{12}{L^{2}}+\frac{l_{G B}^{2}}{2}(R^{2}+\\
-4 R_{\mu \nu} R^{\mu \nu}+R_{\mu \nu \rho \sigma} R^{\mu \nu \rho \sigma})], \end{array}
\end{equation}
where the scale $l_{GB}^{2}$ of the higher derivative term can be chosen to be set by a cosmological constant, $l_{GB}^{2}=\lambda_{G B} L^{2},$ where $\lambda_{GB}$ is the dimensionless parameter. The coefficients of the curvature-squared terms ensure that the equations of motion following from the action (\ref{GB_action}) are second order in derivatives. In the absence of Ostrogradsky instability and other difficulties induced by higher derivatives, Gauss-Bonnet gravity can be a very useful theoretical tool for studying non-perturbative effects of higher-derivative couplings. However, as pointed out in \cite{deBoer:2009pn} for $\lambda_{GB}$ outside of a certain interval, the dual theory suffers from pathologies associated with superluminal propagation of high momentum modes. This would imply that Gauss-Bonnet gravity should loose their privileged non-perturbative status and be treated as any other theory with higher derivative terms, i.e. the coupling $\lambda_{GB}$ should be seen as an infinitesimally small parameter. Here in principle we will not constrain $\lambda_{GB}$ beyond its natural domain $\lambda_{GB} \in(-\infty, 1 / 4]$, limited by the existence of the black brane solution. 

Varying the action (\ref{GB_action}) the Einstein field equations can be derived, this has a pure AdS solution in the absence of any matter sources. We note here that we did not include in the bulk action a contribution from the gauge field, which means that we are not taking backreaction from the flavour branes. To examine holographic superconductivity, we allow for a bulk black-brane metric in order to have a system at finite temperature: 
\begin{equation} \label{GB_metric}
d s^{2}=-a\,f(r) d t^{2}+\frac{1}{f(r)} d r^{2}+\frac{r^{2}}{L^{2}}\left(d x^{2}+d y^{2}+d z^{2}\right)
\end{equation}
dual to a thermal state of a boundary CFT. Here the radial coordinate is $r$ where the boundary theory is located at infinity. The blackening function is given by
\begin{equation} \label{f(r)}
f(r)=\frac{r^{2}}{L^{2}} \frac{1}{2 \lambda_{G B}}\left[1-\sqrt{1-4 \lambda_{G B}\left(1-\frac{r_{H}^{4}}{r^{4}}\right)}\right],
\end{equation}
and the constant $a$ can be chosen to normalize the speed of light at the boundary to be one as
\begin{equation} \label{ac}
a=\frac{1}{2}(1+\sqrt{1-4 \lambda_{G B}}).
\end{equation}
The Hawking temperature is given by
\begin{equation} \label{Tem}
T=\frac{\sqrt{a} r_{H}}{\pi L^{2}}.
\end{equation}
From now on we set the AdS radius $L$ to 1. Note that the model does not work at zero temperature.

As in \cite{BitaghsirFadafan:2018iqr}, we describe the dimension 3 quark bilinear by a scalar field $\psi_i$ where $i$ is one the lines of colour flavour matrix, and baryon number $B=\frac{2}{3}$. One also needs the gauge field associated with $U(1)_B$ where its $A_t$ component describes the chemical potential. We use a Lagrangian given by
\begin{equation}\label{csc_lagrangian}
\mathcal{L}_{CSC}=-\frac{1}{4} F^{\mu \nu} F_{\mu \nu}-\left|\partial \psi_{i}-i BG A \psi_{i}\right|^{2}+3 \psi_{i}^{2}.
\end{equation}
We will define $G$ in detail in section IV, until then we will take it as a constant number. This Lagrangian yields equations of motion
\begin{equation}\label{eom_psi}
\psi_{i}^{\prime \prime}+\left(\frac{f^{\prime}}{f}+\frac{5}{r}\right) \psi_{i}^{\prime}+\frac{B^{2}G^2}{ar^{4} f^{2}} A_{t}^{2} \psi_{i}+\frac{3}{r^{2} f} \psi_{i}=0,
 \end{equation}
and
\begin{equation}\label{eom_At}
A_{t}^{\prime \prime}+\frac{3}{r} A_{t}^{\prime}-\sum_{i} \frac{2 B^{2}G^2}{r^{2} f} \psi_{i}^{2} A_{t}=0.
 \end{equation}

For regularity, one requires that at the horizon  $A_{t}=0$ then from the first equation of motion one finds that $\psi_{i}^{\prime}=-\frac{3}{4 r_{H}} \psi_{i}$. We can always find solution to equations (\ref{eom_psi}) and (\ref{eom_At}) as
\begin{equation}
\psi_{i}=0, \quad A_{t}=\mu-\frac{\mu r_{H}^{2}}{r^{2}}.
\end{equation}
\begin{figure}[h]  
	\includegraphics[width=8cm]{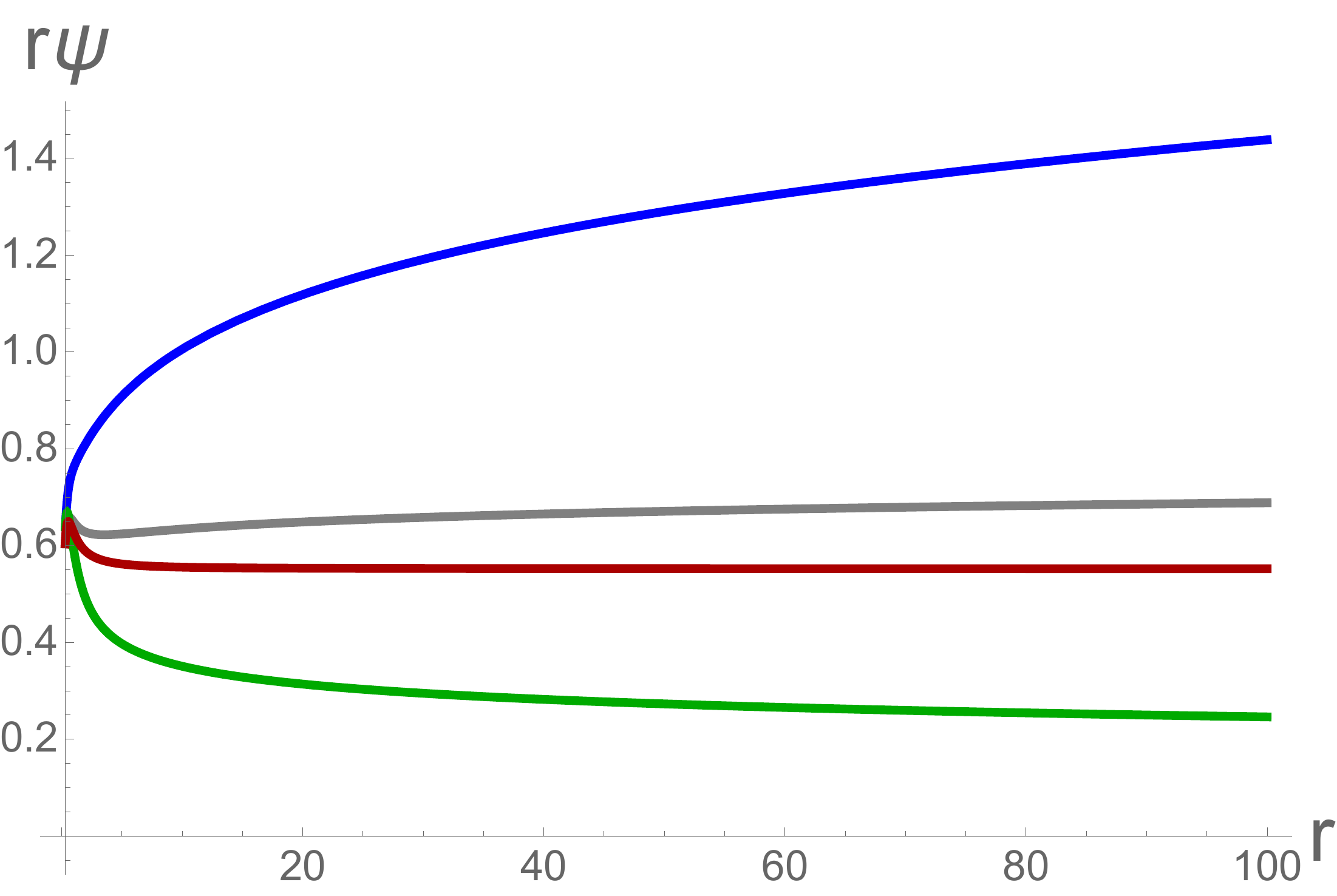}
	
	(a)
	
	\includegraphics[width=8cm]{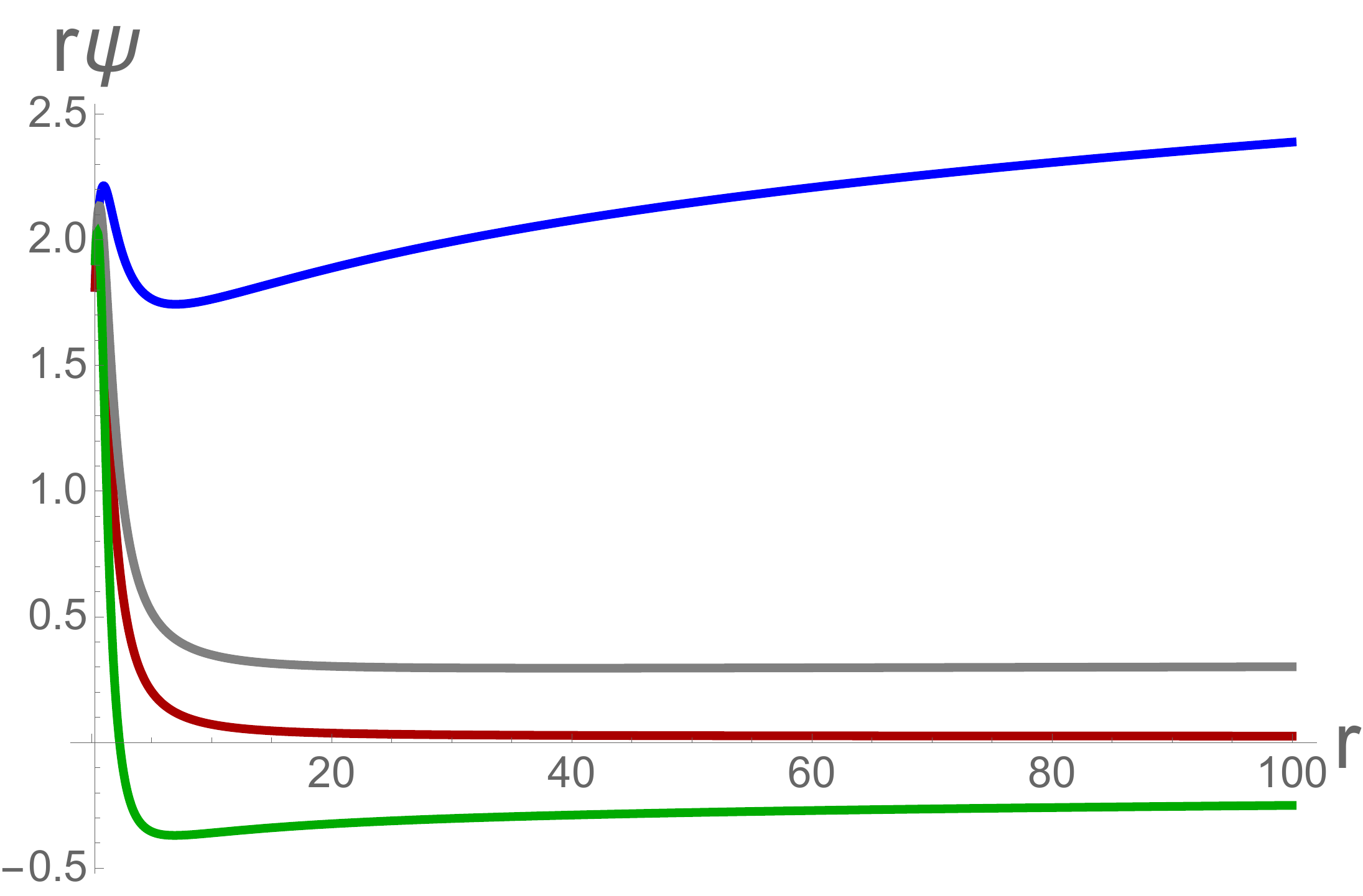}
	
	(b)
	
	\caption{\footnotesize{ \textit{ (a) The $\psi$ functions in the unbroken phase at $T =
				0.1$, $\mu = 1.0$ and initial condition $\psi(r_H)=2$ . (b) The $\psi$ functions in the broken phase at $T =0.1$, $\mu = 5.0$ and initial condition $\psi(r_H)=6$. We varied the values of $\lambda_{GB}$ in both images as $\lambda_{GB}=0.1$ (blue), $\lambda_{GB}=0.025$ (gray), $\lambda_{GB}=0$ (dark red) and $\lambda_{GB}=-0.1$ (green)}}}
	\label{fig:1}
\end{figure}

But more complex solutions can be found numerically by shooting out from the horizon. They take the following form in the UV:
\begin{equation}
\psi_{i}=\frac{J_{c}}{r}+\frac{c}{r^{3}}+\ldots, \quad A_{t}=\mu+\frac{d}{r^{2}}+\ldots
\end{equation}
where from the AdS/CFT correspondence $c$ is interpreted as the Cooper pair condensate $\psi \psi$, which carries both flavour and colour indices. Also $J_{c}$ is the source for $\psi \psi$, and $\mu$
and $d$ are the chemical potential and the density, respectively. There are two constraints and four initial conditions $\psi, \psi^{\prime}, A_{t}, A_{t}^{\prime}$ at the horizon, then one gets a two parameter family of solutions in the IR by fixing  $\psi\left(r_{H}\right)$ and $A_{t}^{\prime}\left(r_{H}\right)$. We label the solutions by values of $\mu$ and $J_{c}$ predicting $d$ and $c$.
\begin{figure}[h]  
	\includegraphics[width=8cm]{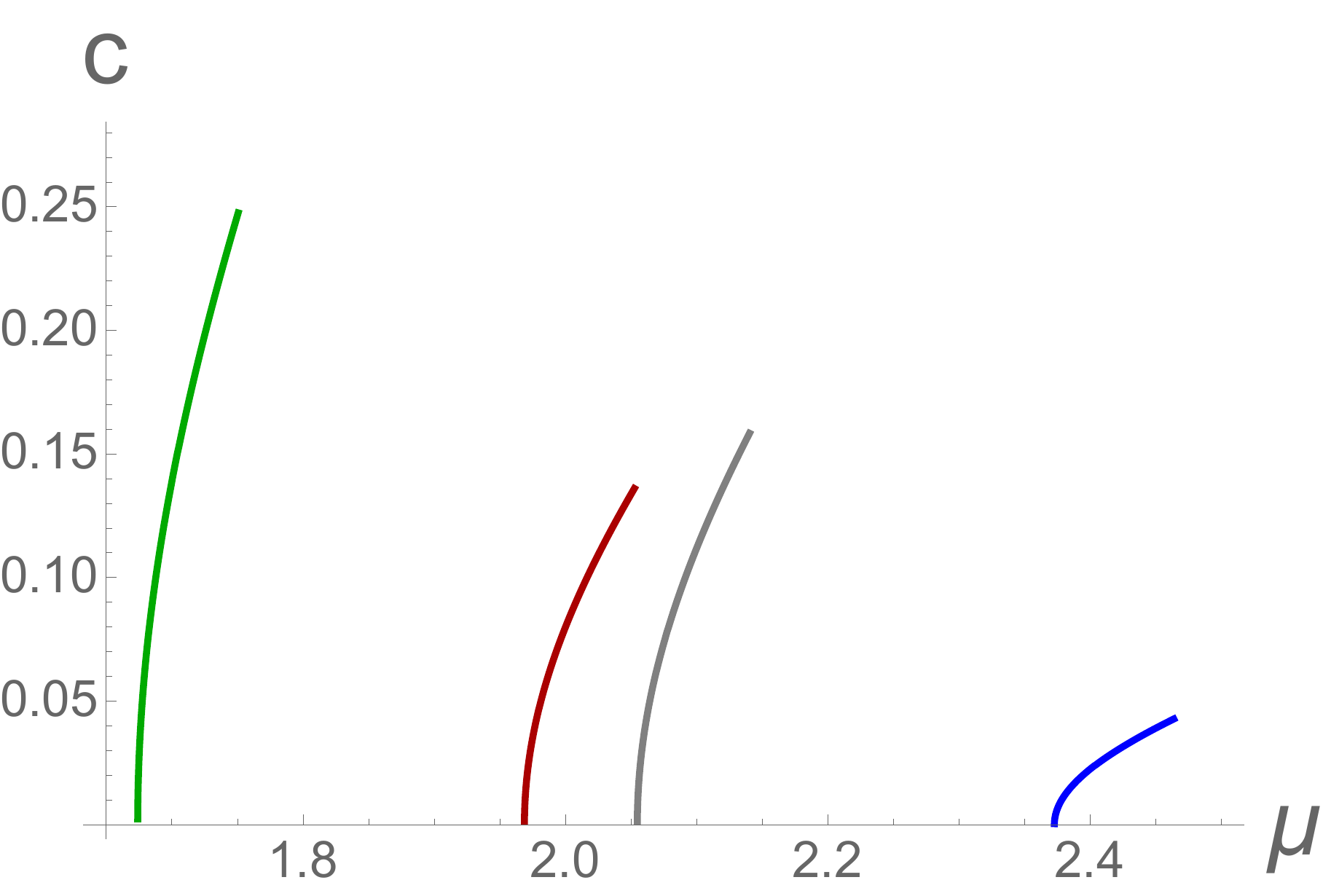}
	
	(a)
	
	\includegraphics[width=8cm]{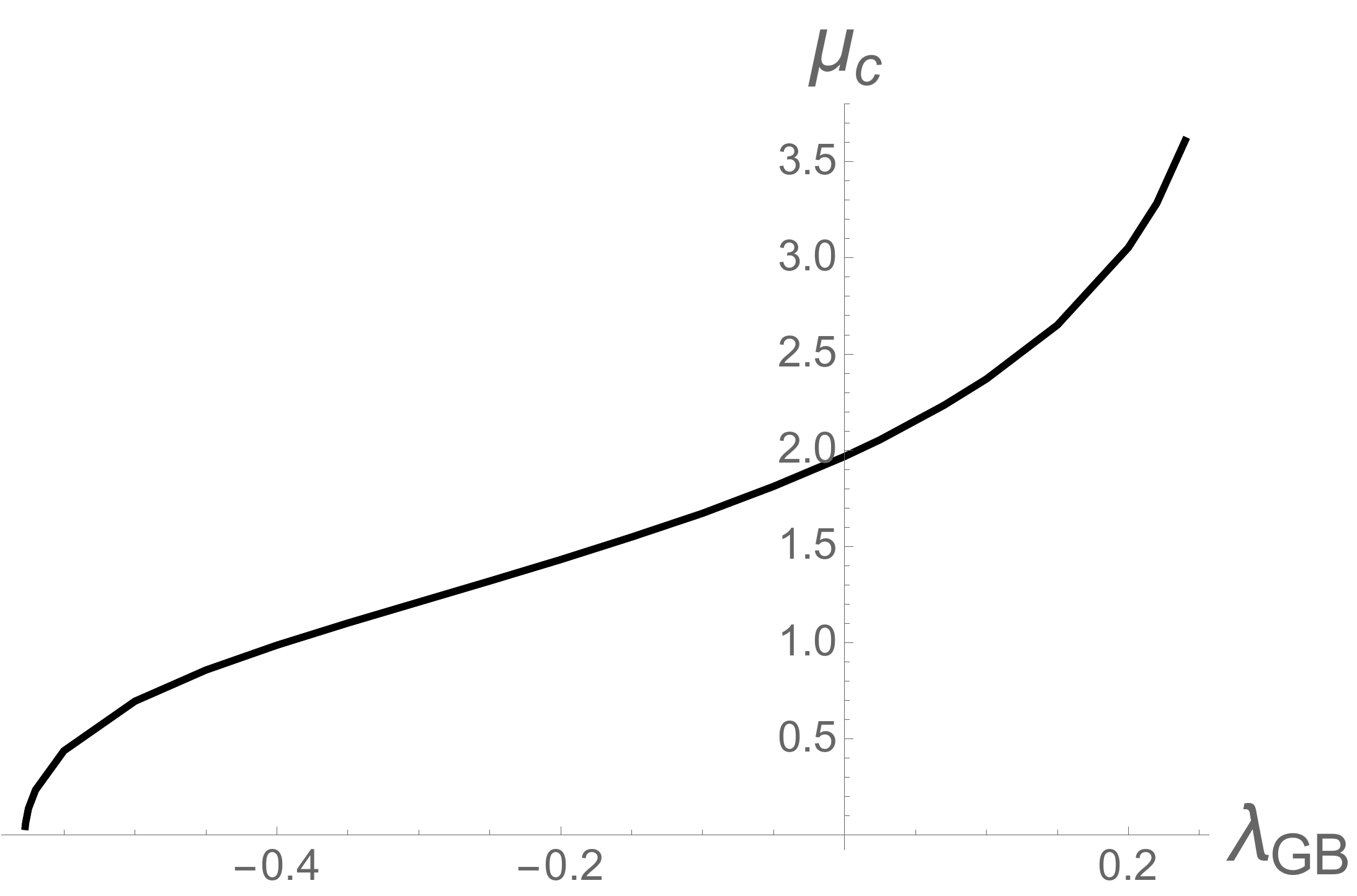}
	
	(b)
	
	\caption{\footnotesize{ \textit{ (a) The condensation vs $\mu$ in the broken phase at $T =
				0.1$ for different values of $\lambda_{GB}$; $\lambda_{GB}=0.1$ (blue), $\lambda_{GB}=0.025$ (gray), $\lambda_{GB}=0$ (dark red) and $\lambda_{GB}=-0.1$ (green). (b) The critical chemical potential $\mu_c$ which is the vaue of $\mu$ at which condensation triggers vs $\lambda_{GB}$}}}
	\label{fig:2}
\end{figure}

In Fig. \ref{fig:1}, we show solutions for the scalar field $\psi(r)$ at fixed temperature, $T=0.1$ and $G=1$. We plot $r \psi(r)$ which asymptotes to $J_{c}$ in the UV. We fixe the initial condition $\psi(r_H)=2$ for all curves in in Fig. 1a, and  $\psi(r_H)=6$ for all curves in in Fig. 1b. We have also adjusted the initial condition for the gauge field $A_{t}^{\prime}\left(r_{H}\right)$ in each case to set $\mu=1.0$ in Fig. 1a, and $\mu=5.0$, in Fig. 1b. Each curve represents a different value of $\lambda_{GB}$ so that from top to bottom $\lambda_{GB}=0.1,0.025, 0, -0.1$.  We observe that by changing  $\lambda_{GB}$ the qualitative behaviour is the same as in \cite{BitaghsirFadafan:2018iqr}. The only exception is the case $\lambda_{GB}=0.1$  (blue curve). One possible reason is because in this case we are above the upper bound $\lambda_{GB}\leq 9/100$ mentioned in \cite{Brigante:2007nu,Brigante:2008gz}, which means that in the boundary theory there is the possibility of superluminal propagation of disturbances of the stress tensor. In \cite{Brigante:2008gz} the authors explore whether a bulk graviton cone behaviour can lead to boundary causality violation by studying the behaviour of graviton null geodesics in the effective geometry. There, they see causality violation for $\lambda_{GB}$ larger than $9/100$.

As one expects, at low $\mu$ there is no symmetry breaking, for  the different values of $\lambda_{GB}$ the only possible solution with $J_{c}=0$ is the trivial one $\psi=0$, therefore $c=0$ (see Fig. 1a). On the other hand, for a higher value of $\mu,$ there is a non trivial solution that asymptotes to $J_{c}=0$ and shows symmetry breaking, since this solution has a non-zero condensate $c$. In Fig. 1b we only show the $\lambda_{GB}=0$ solution showing this but for the other values of $\lambda_{GB}$ similar solutions exist for a different value of the initial condition  $\psi(r_H)$. The physical interpretation in AdS is that the chemical potential generates an effective negative mass squared for the scalar $\psi$ and when it violates the Brietenlohner-Freedman (BF) bound of $m^{2}=-4$ \cite{Breitenlohner:1982jf}  an instability to $\psi$ condensation results. 

We noted that as we increase  $\lambda_{GB}$ above zero one needs to increase the value of  $\psi$ at the horizon to obtain a solution that asymptotes to $J_{c}=0$ compared with the case with $\lambda_{GB}=0$; conversely as we decrease $\lambda_{GB}$ below zero a smaller value of the initial condition for $\psi$ at the horizon is needed  to obtain a solution that asymptotes to $J_{c}=0$ again compared with the $\lambda_{GB}=0$ case. 

From the equation of motion (\ref{eom_psi}) for $\psi$ we can see that the coupling to the gauge potential suppress the mass. Thus we can obtain the effective mass $m_\text{eff}$ and see how it changes with respect to  $\lambda_{GB}$ as
\begin{equation}
m^2_\text{eff}=m^2-\Delta m^2,\:\: \:\:\Delta m^2\equiv \frac{B^2G^2 A^2_t(r)}{ar^2f(r)},
\end{equation}
where its relation to the BF bound has been recently studied in \cite{Nam:2021qwv}.

In Fig. \ref{fig:2}, we plot the condensate as a  function of the chemical potential to see at which value of $\mu$ the condensation is triggered. We do this for different values of $\lambda_{GB}$, which allows us to observe the behaviour of the critical chemical potential $\mu_c$ as a function of $\lambda_{GB}$ for the scalar mass $m^2 =-3$ (see Fig. 2b). We see that as $\lambda_{GB}$ decreases, the value of $\mu_c$ also diminishes, until the value $\lambda_{GB}=-0.78$ where the critical chemical potential approaches zero. Thus, for a sufficiently negative value of  $\lambda_{GB}$ the model shows very small condensation close to zero at all positive values of chemical potential. As a result, beyond this value of  $\lambda_{GB}$ it doesn't look there is a second order transition. The relation between strength of the condensation and values of the $\lambda_{GB}$ has been discussed in \cite{Nam:2021qwv} in the presence of the backreaction.
 \begin{figure}[h]  
	\includegraphics[width=8cm]{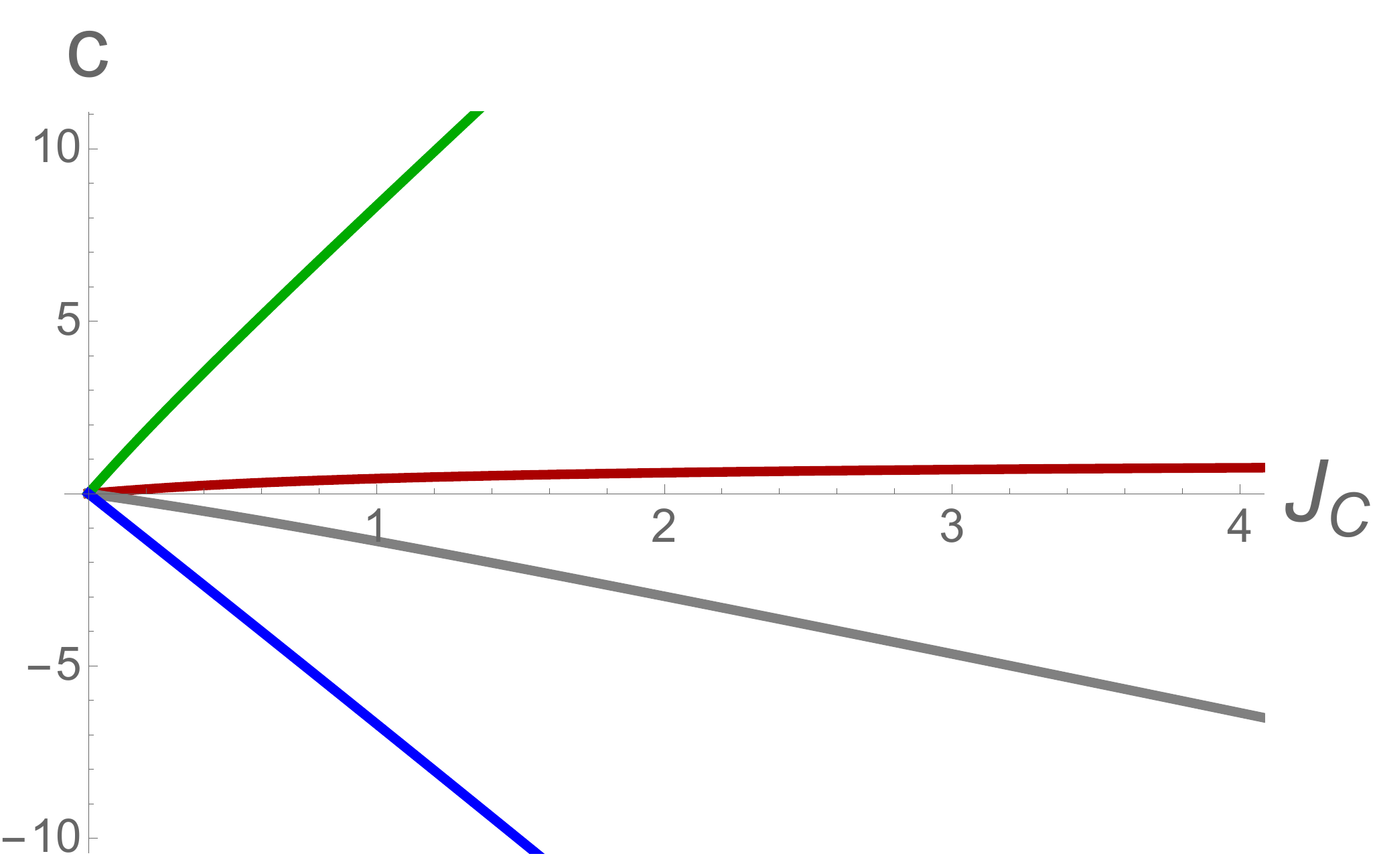}
	
	(a)
	
	\includegraphics[width=8cm]{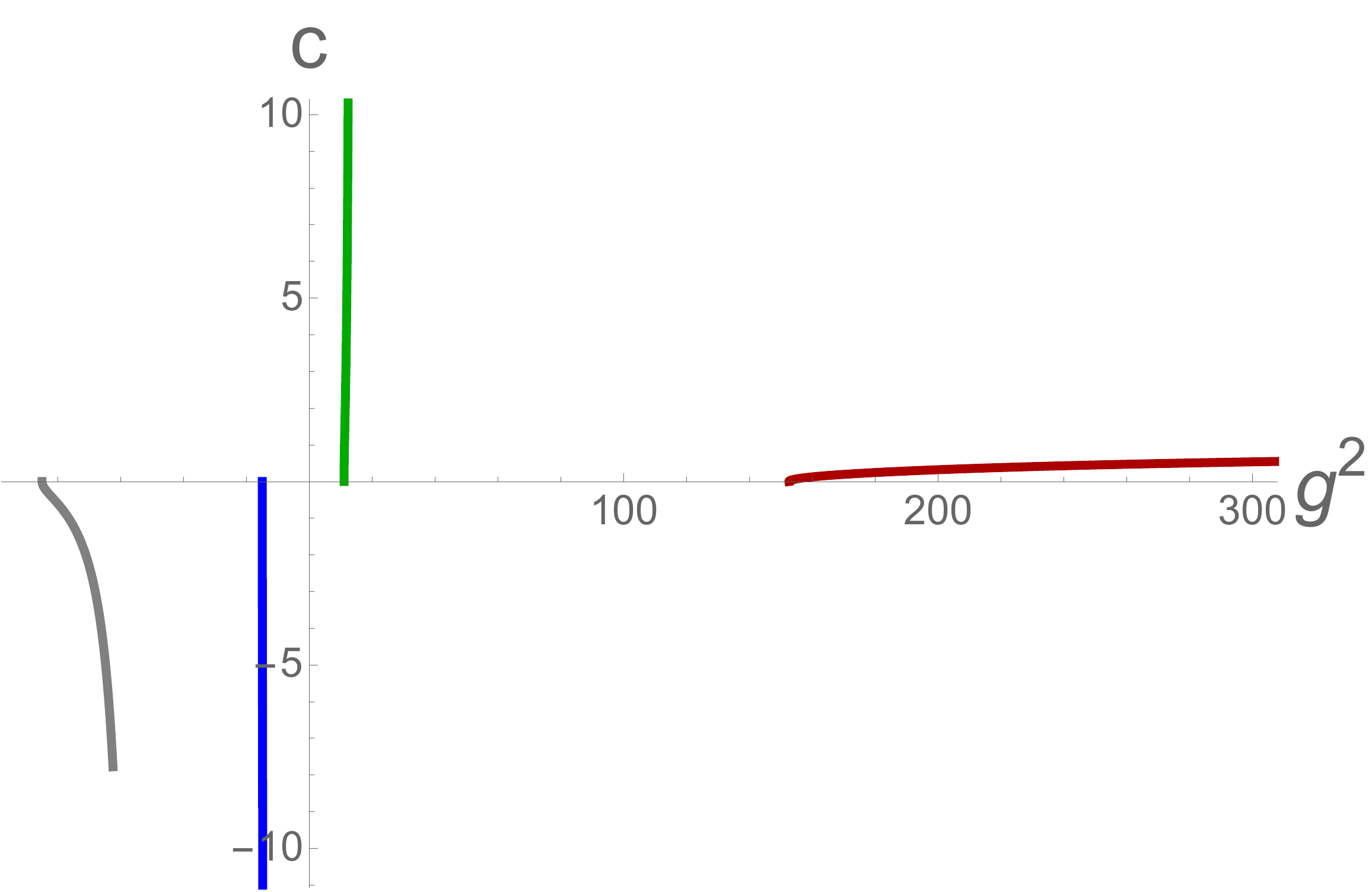}
	
	(b)
	
	\caption{\footnotesize{ \textit{(a) Plot of $c$ against $J_c$ with $\mu<\mu_c$ for embeddings in Fig. 1a.  (b) Plot of $c$ against $g^2$ with $\mu<\mu_c$ for solutions in Fig. 1a. In both images $T=0.1$, $\mu = 0.1$ and we obtained curves for different values of $\lambda_{GB}$; $\lambda_{GB}=0.1$ (blue), $\lambda_{GB}=0.025$ (gray), $\lambda_{GB}=0$ (dark red) and $\lambda_{GB}=-0.1$ (green).}}}
	\label{fig:3}
\end{figure}

Studying the solutions more in detail we see that when $\lambda_{GB}<-0.78$ there still are $\psi$ solutions that break symmetry since they have a positive condensate but the correspondent source $J_c$ is zero, however we can not interpret this in the sense of a second order transition since the value of the condensate and the value of the chemical potential are not correlated as in the cases showed in Fig. 2a. As \cite{Nam:2021qwv}, we find that large magnitude of $\lambda_{GB}$ makes the condensation easier.
\section{Adding NJL interactions}
In this section we attempt to introduce the QCD interactions into the CSC model. In this way, we could think in four fermion operators, the relativistic Nambu-Jona-Lasinio (NJL), as a possible option to model the QCD attraction between quarks, because we assumed that the gluons have acquired a large mass. These operators are an example of a double trace operator and they can be incorporated by using Witten's prescription in \cite{Witten:2001ua}. NJL operators have been studied in holographic superconductors in \cite{Faulkner:2010gj} and in QCD as  \cite{evanskim}.

One finds more in detail how to introduce the four fermion operator into the model in \cite{BitaghsirFadafan:2018iqr}. In short, we consider an operator/source pair such as ${\cal O}, J$ which is described holographically by a field $\psi(r)$ in the $AdS$ background with the following action
\beq S = - \int dr {1 \over 2} \left( r^5 (\partial_r \psi)^2  - 3 r^2 \psi^2 \right). \label{psiaction}\eeq
The solutions of the equations of motion in the UV take the form
\beq \psi = J/r + {\cal O}/r^3. \label{sol} \eeq
Based on Witten's prescription, one should add a UV surface term into the action at the cut off $\Lambda$. We want to include in the boundary field theory a term of the form
\beq \Delta {\cal L} = - {g^2 \over \Lambda^2} {\cal O} {\cal O}, \label{nine} \eeq
where ${\cal O} \neq$ 0, then this term generates a source  
\begin{equation} J \simeq {g^2 \over \Lambda^2} {\cal O}. \label{condition} \end{equation}
This  is the main condition that we apply to the solutions of the equation of motion we already have. This has been done in Fig. \ref{fig:3} for the holographic color superconductor model in the unbroken phase ($\mu<\mu_c$). Here solutions with $J_c \neq 0$ are interpreted as having zero intrinsic $J_c$, but, in the presence of NJL interactions. Then the four fermion interaction would generate the source $J_c$ in the UV.

To explain this in detail, consider  for example the holographic field solutions $\psi(r)$ in Fig. 1a, as we discussed in the previous section, there are two sets of solutions depending on the critical value of the chemical potential $\mu_c$. For solutions below $\mu_c$, we plot the condensate $c$ against the NJL coupling $g^2$ in Figure 3b using equation (\ref{condition}). Here we have taken the UV cut off $\Lambda=10$ numerically. One finds that a critical value of the NJL coupling $g^2$, that triggers symmetry breaking at a second order phase transition. For the cases of $\lambda_{GB}=0$ and $\lambda_{GB}=0.1$ we see that the value of $g^2$ at which condensation turns on is positive, whereas for positive values of $\lambda_{GB}$ the critical value of $g^2$ is negative. For the solutions with $\lambda_{GB}>0$, in the UV, $J_c$ and $c$ have opposite signs, with the condensate being always negative as we can see in Fig. 3a. On the other hand for $\lambda_{GB}\leq0$ the condensate is positive and we recover the same behaviour as in reference \cite{BitaghsirFadafan:2018iqr}.

Similarly we could translate the functions of Figure 1b to  show $c$ vs $g^2$ but as in the previously reported case with $\lambda_{GB}=0$,  even at $g^2=0$ there is already symmetry breaking. At a value of chemical potential $\mu>\mu_c$ the two interesting additional features showed for the $\lambda_{GB}=0$ case are still present. Firstly there are solutions at negative and repulsive, $g^2$. Which is reasonable because there is a symmetry breaking at zero NJL coupling and switching on a repulsive NJL coupling reduces the condensation. The second interesting feature is that the condensation can only switch off at infinite repulsive four fermion interaction.  The remaining structure in the $c-g^2$ plane is the translation of the spiral in the $c-J_c$ plane seen previously in \cite{BitaghsirFadafan:2018iqr}. Thus one may conclude that in the presence of higher derivative corrections the above two interesting features still present and the intrinsic attractive interaction is more complex in this holographic model. Then it would be worth to explore other possibilities such as different gravity backgrounds to study these features in the model.

\section{The phase diagram at finite coupling}
Now we can describe the effect of the Gauss-Bonnet correction in the bulk to the colour superconducting phase of QCD. The phase diagram is described in $T-\mu$ plane, chiral symmetry breaking and confinement scales have been recently studied in \cite{Evans:2020ztq}. Here we assume that the chiral phase transition occurs at $T^2 + \mu^2 = \Lambda_c^2$. We numerically set $\Lambda_c=1$ then the UV cut off of the holographic model is $\Lambda=10 \Lambda_c$ where we read off condensation $c$ and source $J_c$.  We also assume a phase with a $q\bar{q}$ condensate below $\Lambda_c$. As it was mentioned even in the presence of the Gauss-Bonnet coupling corrections, NJL interactions can not switch off the condensation at each $T$ and $\mu$. This confirms that intrinsic QCD interactions are more subtle than the four fermion interaction. 
\begin{figure}[]
	\centering
	{\includegraphics[width=7.5cm]{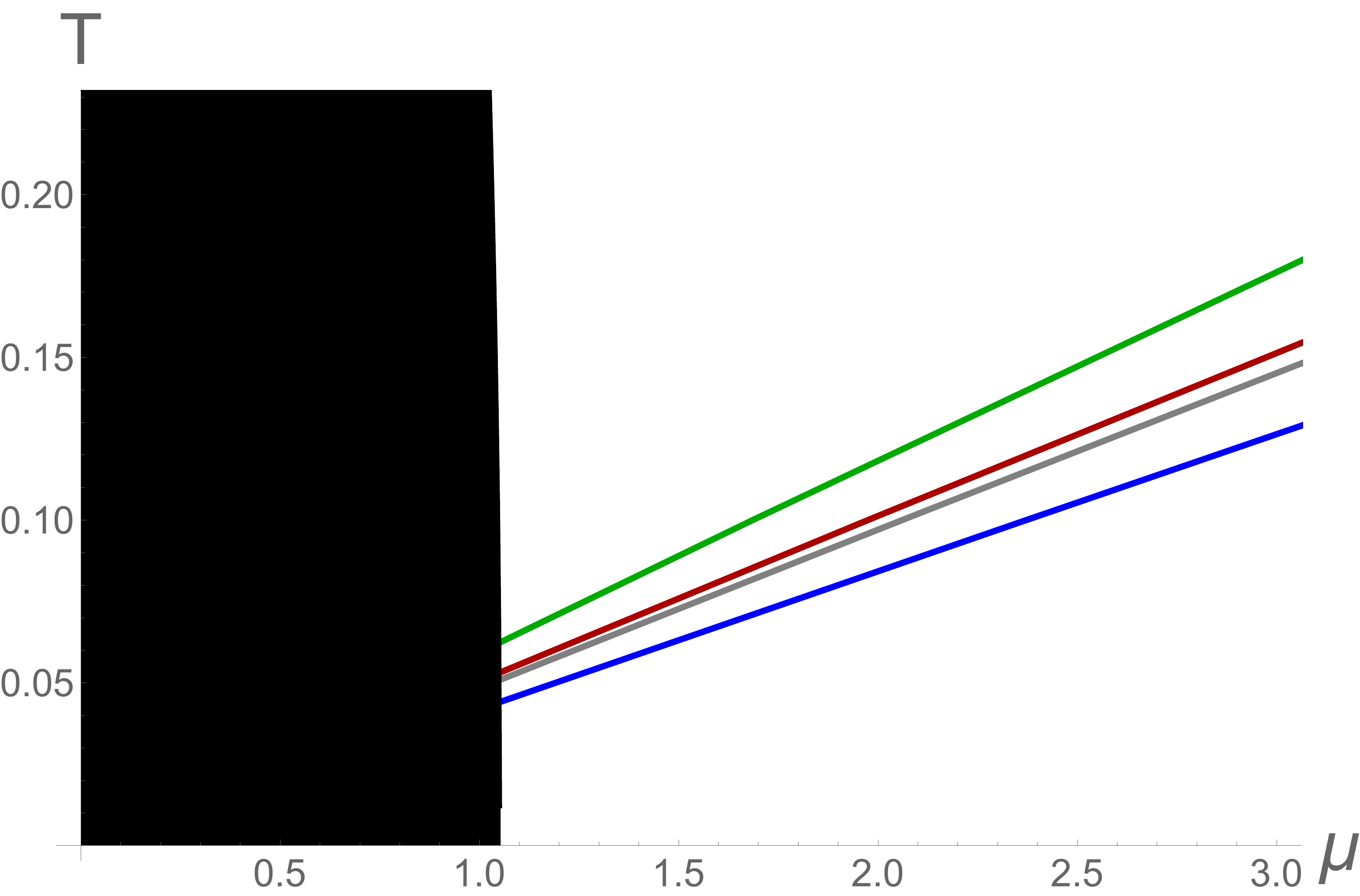}   }
	\caption{Plot of the superconducting phase boundary at $G=1$  and different $\lambda_{GB}$ in the T - $\mu$ plane; $\lambda_{GB}=0.1$ (blue), $\lambda_{GB}=0.025$ (gray), $\lambda_{GB}=0$ (dark red) and $\lambda_{GB}=-0.1$ (green). The black region is expected to be the chirally broken phase below a scale of $\mu^2+T^2 = 1$. }
	\label{fig:4}
\end{figure}
As it was proposed in \cite{BitaghsirFadafan:2018iqr}, there is a way to affect the condensation by modifying the strength of the interaction between $A_t$ and colour pairing $\psi$. Using the main assumption for the global symmetry of the colour of quarks, one reads the action \eqref{csc_lagrangian} where the interaction term of $A \psi$ gives information about the broken phase. Here $G$ is a new coupling which is related to the one loop running coupling as 
\beq G^2 = {\kappa \over b \ln (T^2 + \mu^2)/\Lambda_c^2}, \hspace{1cm} b =  11 N_c/3 -  2 N_f/3. \label{qcdG}\eeq

The parameter $\kappa$  is a free parameter and it reflects the strength of the attraction that generates the $\bar{3}$ of the colour condensate, see \cite{BitaghsirFadafan:2018iqr}. To study the effect of $\lambda_{GB}\neq0$ in the  phase boundary for the superconducting phase, first for each value of $T$ and $G$ we obtain the critical chemical potential and then plot $\mu_c(T)$. This is shown in Figure 4. In this figure, the black region corresponds to chirally broken phase and we have fixed $\Lambda_c=1$. One can summarize the main results as follows:
\begin{itemize}
	\item As $\lambda_{GB}$ grows the temperature at which the boundary phase is present decreases significantly.
	\item For $\lambda_{GB}<0$ the temperature at which the boundary phase is present is larger than in the $\lambda_{GB}=0$ case.
\end{itemize}

Then we can control the boundary phase by tuning $\lambda_{GB}$ which is very useful because one can study smaller temperatures. However, there is a bound on the maximum positive value of the coupling which is $\lambda_{GB}=0.24$.  

The other free parameter of the model $\kappa$ comes from constraints by studying one gluon exchange interactions of $qq$ and $q\bar q$. It was argued in \cite{BitaghsirFadafan:2018iqr} that a sensible range of it is between 1 and 20 where the color superconductivity occurs. As (\ref{qcdG}), it expresses the the intrinsic interaction between the fields $A_t$ and $\psi$.

Now using these tools one can construct the phase diagram from the analysis of Figure 4. We fix $N_f=3$ and for each value of $G$ from (\ref{qcdG}), plot the circles in the $T-\mu$ plane. There are crossing points with corresponding curves with the sames $G$ value where we identified them in figure 5. This figure shows the phase diagram in the presence of finite coupling corrections. Here $\kappa= 10$ and from to to bottom $\lambda_{GB}=-0.1,0,0.025, 0.1$. As it is clearly seen,
\begin{figure}[]
	\centering
	{\includegraphics[width=8cm]{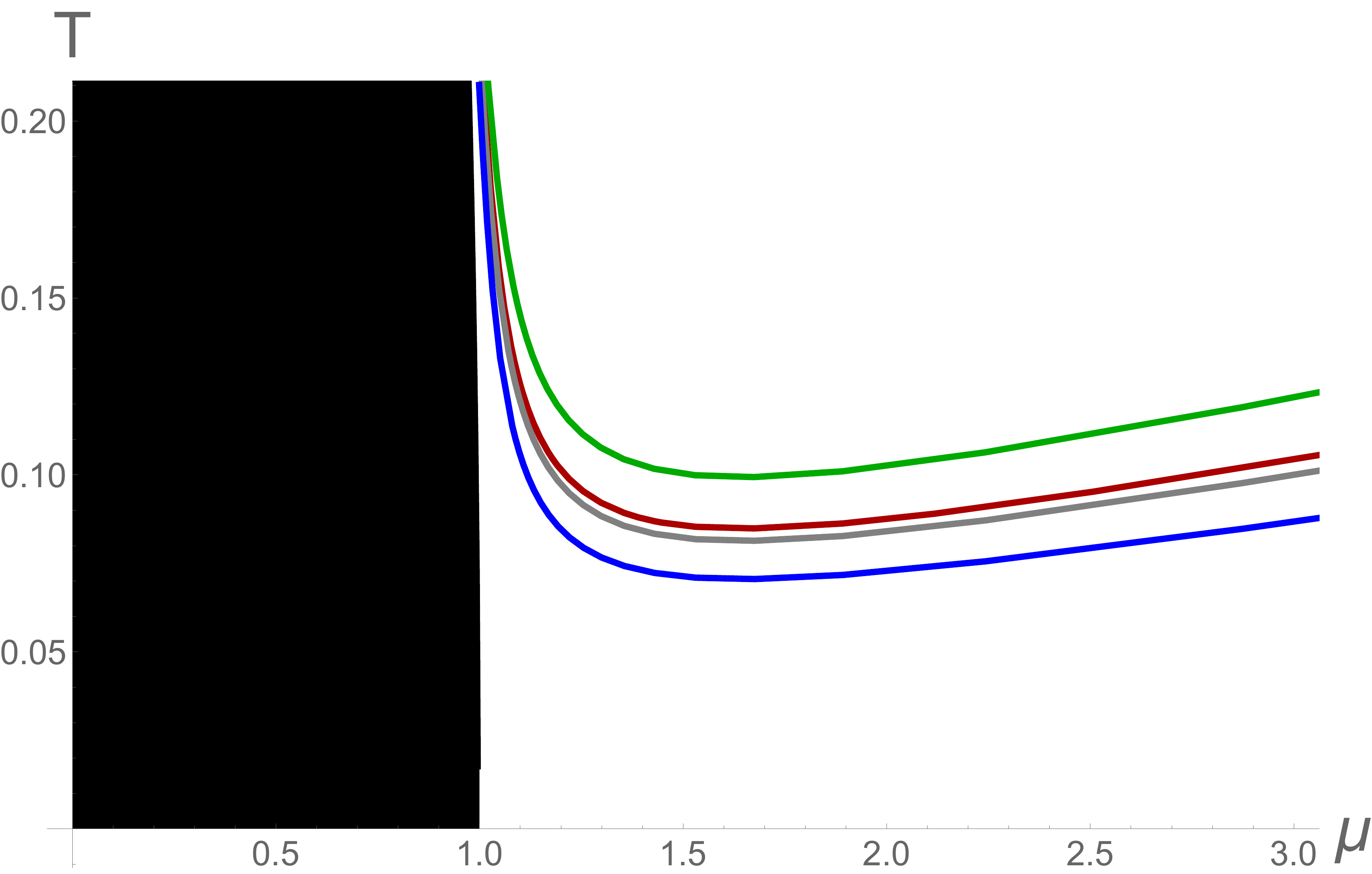}   }
	\caption{QCD phase diagram for different values of $\lambda_{GB}$; in the blacked out area  chiral symmetry breaking is expected. The remaining phase edges shows where the CFL phase is present for the choices of $\kappa=10$ and $\lambda_{GB}=0.1$ (blue), $\lambda_{GB}=0.025$ (gray), $\lambda_{GB}=0$ (dark red) and $\lambda_{GB}=-0.1$ (green).  }
	\label{fig:mass}
\end{figure}
the colour superconductor phase depends on $\lambda_{GB}$, however the shape of the phase curves do not change. For example, very close to $\Lambda_c=1$ the QCD coupling gets very strong and the colour superconducting phase hugs the phase boundary up to higher values of the temperature. One finds that 
\begin{itemize}
	\item For positive values of $\lambda_{GB}$, we observe that the phase boundary moves to lower values of $T$ at a given $\mu$.
	\item For negative values of $\lambda_{GB}$ a higher temperature needs to be reached in order to get condensation (green curve in Fig. 5).
\end{itemize} 
By considering Gauss-Bonnet corrections, still we are in the sensible regime of the colour superconducting phase. Because, this phase is predicted to exist when temperature is bellow of 0.15 $\Lambda_c$ \cite{Alford:2007xm}. For fixed $\kappa \simeq 10$ and the expected chiral transition temperature $\Lambda_c \simeq 175$ MeV, the estimate might be 20 MeV where can be changed a few percent by adding Gauss-Bonnet correction as we see in figure 5.
\section{Discussion}
In this paper we followed the holographic bottom up model of colour superconductivity proposed in \cite{BitaghsirFadafan:2018iqr} by considering higher derivative corrections in the bulk. We have studied Gauss-Bonnet gravity which, from the AdS/CFT correspondence, is related to finite coupling corrections in the boundary theory. The charged scalar field $\psi$ in the bulk is describing diquark condensation, while the U(1) gauge coupling should be interpreted as baryon number. First we studied unbroken and broken phases and how the critical chemical potential, which separates both phases, changes with higher derivative corrections. This is shown in Figure 1 and Figure 2. It was shown that a negative value of $\lambda_{GB}$ can significantly reduces the critical value of the chemical potential where Cooper condensation starts to form. Next, we introduced NJL operators as the responsible operators for the four fermion interaction between quarks and studied the changes that the Gauss-Bonnet corrections induce in the quark condensate in Figure 3. We found that finite coupling corrections do not change the main features of the model and concluded that intrinsic attractive interaction can not be switched off by such corrections.  We then dropped NJL operators and just tuned the coupling between the scalar and gauge fields in the gravitational theory to relate it with the QCD  running coupling. We found the phase edge for the colour superconductivity phase as a function of temperature and chemical potential. We also studied how the phase diagram changes with  Gauss-Bonnet corrections. It was shown that by tuning the $\lambda_{GB}$ one can study smaller temperatures in the $T-\mu$ phase diagram. This is the interesting regime for studying physics of compact stars, then it would be desirable to find finite coupling corrections for the holographic description of neutron stars.

\noindent {\bf Acknowledgements:}
We would like to thank N. Evans for reading the manuscript and his interesting comments.

\end{document}